\documentclass[12pt]{article}
\usepackage{epsfig}
\usepackage{psfrag}
\usepackage{amsmath}
\usepackage{lscape}
\usepackage{multirow}
\usepackage[super, comma, sort&compress]{natbib}
\begin{document}

\title{Automated sampling assessment for molecular simulations using the 
effective sample size}

\author{Xin Zhang, Divesh Bhatt, and Daniel M. Zuckerman}

\maketitle
\begin{abstract}
To quantify the progress in development of algorithms and
forcefields used in molecular simulations, 
a method for the assessment of the sampling quality
is needed.  We propose a general method to 
assess the sampling quality through the estimation of the number
of independent samples obtained from molecular simulations.
This method is applicable to both dynamic
and nondynamic methods and utilizes the variance in the populations
of physical states to determine the ESS. We test the correctness and
robustness of our procedure in a variety of systems -- two--state toy
model, all--atom butane, coarse--grained calmodulin, all--atom
dileucine and Met--enkaphalin. 

We also introduce an automated procedure to
obtain approximate physical states from dynamic trajectories: this procedure
allows for sample--size estimation for systems for which physical states
are not known in advance.

\end{abstract}

\section{Introduction}

 The field of molecular simulations has expanded rapidly in the last
two decades and continues to do so with the advent of progressively
faster computers. Due to this, significant effort is put in the
development of more sophisticated algorithms\cite{frenkel_smit} 
and forcefields for use in both physical and biological sciences.
To quantify progress, it is critical to answer questions about the
efficiency of the algorithms and the accuracy of the forcefields.
This is especially true for large biomolecules that are slow to
sample. Such questions demand a gauge to assess the quality of the 
generated ensembles that is applicable regardless of the type of
method used to generate the ensembles.

 Ensembles are of fundamental importance in statistical mechanical
description of physical systems: thermodynamic properties are obtained
from these ensembles. In addition to the accuracy, the quality of
ensembles is governed by the amount of information present in the
ensemble. Due to significant correlations between successive
frames in, say, a dynamic trajectory, the amount of information
does not equate to the total number of frames. Rather, the number
of independent samples in the ensemble (or the effective sample size) 
is the required gauge for the sampling quality.
This effective sample size (ESS) has remained difficult to assess
in general. In this work, we present an efficient method to determine the 
ESS of an ensemble -- regardless of the method used to generate the
ensemble -- by quantifying the variances in physical states of the system
under consideration.

A conventional view of sample size is given by the following equation:
\begin{equation}
N=\frac{t_{\mathrm{sim}}}{t_{\mathrm{corr}}}
\label{e1}
\end{equation}
 where $t_{\mathrm{sim}}$ is the simulation time, 
and $t_{\mathrm{corr}}$ is the largest correlation
time in the system.
Thus, significant effort has been invested in developing
methods to calculate the correlation time.  However, the estimation
of the correlation time needs the computation of a correlation function, 
which requires a significant amount of data.
Other assessment approaches have, therefore,
been proposed.  Mountain and Thirumalai\cite{ergodic1,ergodic2} 
introduced the ``ergodic
measure'', where the time required for the observable to appear
ergodic is decreased. Flybjerg and Petersen\cite{block} developed a block 
averaging method with the idea that with an increase in size of blocks, 
adjacent blocks are less correlated.

Most of these methods analyze particular observable, however,
different observable can give different correlation times. 
For example, in a typical model, bond lengths are decorrelated faster than 
dihedral angles.  However, bond lengths are never fully decoupled from 
the rest of the system: slower motions must ultimately couple to
the fast motions and influence their distributions. Moreover, we are 
not trying to compute a particular ensemble average but to generate a truly 
representative ensemble of configurations. Thus, there is significant
ambiguity in the use of observable to estimate the correlation times.

In order to overcome this ambiguity, the idea of ``structural
histograms'' was proposed.\cite{ed2} In this idea, a dynamic trajectory is
divided into structural bins based on several reference structures.
For correlated samples, the average variance of these structural bins 
depends upon the time increment between frames used to contruct the
structural bins. In the limit of ideal sampling, this variance
becomes independent of this time increment. Accordingly, this method
determines the time increment that results in such an ideal sampling.
For most complex systems, the data are usually limited, leading to
significant uncertaintities in the estimation of the average variances,
thus, this method gives only logarithmically correct (factor of $\approx$2)
results for the ESS.

All the methods discussed above require dynamic trajectories (such as those
obtained by molecular dynamics (MD),\cite{md} 
Metropolis Monte Carlo (MC),\cite{metropolis} or Langevin
dynamics) for analyzing the correlations. As such, these methods are not
applicable to non--dynamic methods, such as Replica Exchange
and its variants\cite{rex1,rex2,rex3} and polymer--growth 
MC.\cite{rosenbluth,wall,grassberger,liu}
Thus, an universal method that can estimate the effective sample size
for both dynamic simulations and non--dynamic methods is particularly 
important -- allowing for, in addition to the sample size estimation for 
ensembles generated via non--dynamic method, comparison of efficiency between
the two classes of methods.

In this work we propose a novel method for the estimation of the effective 
sample size that is applicable universally to ensembles generated using both
dynamic and non--dynamic methods. The main idea in this work is to quantify
the variance of populations in approximate physical states, 
and estimate the effective
sample size from this variance by mapping the problem into a binomial
distribution: either a configuration is in one particular physical state 
or is not. The use of physical states is intuitive: transitions between 
physical states represent the slowest
timescales in a system, and the simulation must be long enough to
show several such transitions for the generated ensemble to be

 This method differs from the structural histograms method in various
crucial ways. First, the use of population variances allows the present
work to estimate the ESS for ensembles generated using both the dynamic 
and non--dynamic methods. Second, the use of physical--state variances
allows us to probe the slowest timescale in the system, and the effective
sample size of an ensemble must be governed by this timescale.

 Accordingly, a important prerequisite for the estimation of ESS
is the determination of physical states. 
In this work, we use a particularly simple method for the determination 
of physical states that uses information present in a dynamic trajectory
regarding the transition rates between different regions -- regions
showing high transition rates with each other belong in the same
physical state. Further, this procedure also highlights the
hierarchical nature of the energy landscape.
Our approach to determine the physical states is based on ideas
of Chodera {\it et al.}\cite{chodera} who
developed an automated procedure to determine approximate physical states
in a system from a dynamic trajectory by determining a division of
the total configuration space that maximizes the self transition
probabilities ({\it i.e.}, the divisions represent metastable states).

 The manuscript is organized as follows. First, we describe in detail the
procedure we use to estimate the effective sample size. Then, we present
results for several models with different levels of complexity
-- a two--state toy model, butane, calmodulin,
di--leucine, and Met--enkaphalin. This is followed by a discussion of
the practical aspects of the procedure. And, finally, we present conclusions.
Further, in the Appendix, we show a simple, automated way to determine
approximate physical states that are required for obtaining reliable
estimates of the effective sample size.

\section{Method}

 State populations are the most fundamental descriptors
of the equilibrium ensemble, and any algorithm (either dynamic or
non dynamic) must sample state populations correctly.
Hence, we claim that variances in populations in ``regions'' of configurational
space form a fundamental basis for estimating the ESS.
We shortly define the most appropriate definitions of ``regions'' and
how to compute the variances, but first, we introduce the main idea relating 
these variances to the ESS.

The key idea that we use in determining the ESS, $N_{\mathrm{eff}}$ is that
the occurrence of a configuration (irrespective of whether generated using
a dynamic or a non--dynamic method) in a region (or bin) of the configuration
space can be mapped onto the binomial distribution: the configuration is
either in bin $j$ with probability $p_j$, or not in that bin with probability
$1-p_j$. The value of $p_j$ equals the fractional bin population that is
formally and uniquely given by the partition function.
In the limit of all such Bernoulli trails being independent,
the variance in the fractional population of bin $j$ is
\begin{equation}
\sigma_j^2(N)=\frac{p_j(1-p_j)}{N}
\label{e2}
\end{equation}
where $N$ is the number of independent trials.

 Equation~\ref{e2} forms the basis for estimating ESS as follows. 
In essence, independent estimates of populations (of bins or,
more accurately, of physical states), $p_i$, are available,
then the computed variances in these populations lead to
estimating $N_{\mathrm{eff}}$ ($\equiv N$) by inverting eq~\ref{e2}.
In this work, we use Voronoi bins: the details of binning procedure,
and the subsequent determination of approximate physical states,
are given in the appendix.

 The above discussion naturally leads to several issues. Firstly, it
is not immediately obvious how to obtain independent estimates of $p_j$ to
compute the bin variances from
typical dynamic and non--dynamic simulations. 
Secondly, what is the best way to define the bins or regions such
that we obtain meaningful estimate of the ESS? Thirdly, it is likely
that different bins give different
values of the ESS. Below, we discuss these issues separately.

\subsection{Population variances from independent probability estimates}

First, we discuss how independent estimates of fractional populations in
the bins can be obtained in both dynamic and non--dynamic simulations.
As discussed in Introduction, non--dynamic methods have no intrinsic
correlation time, unlike dynamic simulations. Thus, the only statistical
analysis possible for both types of simulations is via several independent
simulations. Average fractional bin populations and bin population
variances can be computed from these independent simulations leading to
an estimate of $N_{\mathrm{eff}}$ using eq~\ref{e2}.

To facilitate exploration of important parts of the configuration space,
we generate the independent simulations from different starting configuration.
Further, all the independent simulations should be of the same length
(dynamic simulations) or generate same number of configurations 
(non--dynamic simulations). This requirement is due to our assumption that
all the independent simulations have same $N_{\mathrm{eff}}$.
Alternately, it is possible to perform a very long simulation,
divide it into 10 or 20 equal---length segments, and consider these
segments as independent simulations. However, care should be taken
in this case , as we discuss later.

\subsection{Appropriate division of configuration space: physical states}

 Equation~\ref{e2} can, in principle, be applied to any decomposition
of the configurational space. However, the separation of timescales
in a system -- fast relaxation
within a physical state and slow relaxation between two physical states --
leads to the use of physical states as the most appropriate decomposition
to use with Figure~\ref{e2}.

 Due to a seperation of the timescale, state populations are slow
to converge compared to sampling within a state. However, a prerequisite
for a correct determination of state populations is a proper
sampling within the states.
Thus, an algorithm must sample correctly within each state of interest.
The ratio of partition functions for states $i$ and $j$, with
configurational--space volume $V_i$ and $V_j$, respectively, is
\begin{equation}
\frac{\mathrm{prob}(i)}{\mathrm{prob}(j)}=
 \frac{\mathrm{Z_{i}}}{\mathrm{Z_{j}}}=
 \frac{\int_{V_{i}}d\mathrm{\mathbf{r}\exp(-U(\mathbf{r})/k_{\mathrm{B}}T)}}
  {\int_{V_{j}}d\mathbf{\mathrm{\mathbf{r}}}
 \exp(-U(\mathbf{r})/k_{\mathrm{B}}T)}
\label{e4}
\end{equation}
where $Z_i$ is the partition function for State $i$, $U$ is the
potential energy of the system, $T$ is the temperature, and $\mathbf{r}$
are the configurational--space coordinates.
This ratio is converged only if both the integrals are converged.

 Variances in the physical states probe the sampling quality within
a state, and, thus, are appropriate measures to determine the sampling
quality using eq~\ref{e2}.

Further, bins much smaller than the physical states lead to incorrect estimates
of the sample size using correlated trajectory that is typically
obtained from a molecular simulation. We discuss this in more detail in
Section~\ref{phys} using results from a butane trajectory.

 In the Appendix, we describe in detail a simple and efficient way to determine
physical states in a system. We emphasize that
our procedure to calculate ESS is independent of the procedure used to
determine physical states.  Briefly, we first decompose the total 
configurational space into bins (Voronoi bins, based on reference structures).
Subsequently, we combine bins based on rates between the bins: bin pairs with
high transition rates are expected to be part of one physical state 
({\it i.e.}, devoid of significant internal barriers). This approach
naturally leads to a hierarchical picture of the energy landscape, as we
discuss in more detail in the Appendix. 

\subsection{Multiple estimates of ESS}

Here, we discuss the practical scenario that each physical state gives 
its own estimate of the ESS. Although it is possible
that the sample sizes in different regions are different from each
other, we investigate the slowest timescale, which corresponds
to the highest energies barrier in the systems. In other words, the
sample size we are looking for is the overall sample size in the whole
trajectory. The simulation could sample the space better in one region
than the others, however the region that gives the smallest sample size
reflect sampling quality and is the sample size we use.

 Additionally, taking the slowest time scale has the advantage that
the estimated ESS is insensitive to whether a physical state is determined
inaccurately. Such an ianccurate physical state can arise, for example,
if the bin combination procedure discussed in the Appendix results
in a physical state that straddles a barrier.

\section{Results}

\subsection{Toy two--state system}

 First, we establish the correctness of our method for estimating
$N_{\mathrm{eff}}$. For this purpose, we choose a system in which
all the samples are independent and, hence, we have a prior 
knowledge of the exact 
answer. The system has two ``states'' defined such that first state
consists of numbers below 0.5, and the second state consists of numbers
above 0.5. The sampling consists of independent drawing of random
numbers between 0 and 1, and the sample size is the number of independent
draws, $N$. 

 To determine whether an accurate estimate of $N_{\mathrm{eff}}$
($\equiv N$ in this system) is obtained, we also compute both
the mean value and standard deviation of $N_{\mathrm{eff}}$. As
suggested by eq~\ref{e2}, this required computation of variances
of both the mean population and the population variance (these
quantities are equal across the states for a two--state system).
Further, care must be taken to account for the nonlinear dependence of 
$N_{\mathrm{eff}}$ on the state variance. The details of this
computation are given in the supplementary information.

 For $N=2000$, we obtain a mean value of $N_{\mathrm{eff}}$ as
2004, with a standard deviation of 57.4. Similarly, for $N=4000$,
we obtain a mean $N_{\mathrm{eff}}$ as 4041 with a standard deviation
117.6. Clearly, our basic premise of using the binomial distribution
inherent in eq~\ref{e2} is correct.

\subsection{Systems with {\it a priori} known physical states}

 In contrast to sampling in the toy system discussed above, sampling in
molecular systems is not typically independent, and, thus, the sample size is
not known {\it a priori}. For this purpose we compute $N_{\mathrm{eff}}$
using the procedure outlined above using sampling obtained from a
typical molecular simulation. Furthermore, a knowledge of physical states
allows an independent estimate of the effective sample size: a trajectory
reentering a physical state is expected too be independent of its previous
history in that particular state, and the counting transitions in and out
of a physical state gives an estimate of the effective sample size.
Although physical states determined using the analysis described in the
Appendix can be used to ``count'' transitions, {\it a priori} knowledge
of physical states makes this estimate especially straightforward.

 The particular systems we use are those obtained for an all--atom
butane model and a coarse--grained calmodulin model. The trajectories
for the all--atom butane model are obtained at 298 K using OPLSAA
force field in vacuum with Langevin dynamics (as implemented
in Tinker v. 4.2.2). This system has three well--known physical
states: trans, gauche+, and gauche-. For calmodulin, the trajectories
were generated by employing Monte Carlo simulations using a backbone 
alpha--carbon double--G$\bar{\mathrm{o}}$ model stabilizing the two
known physical states (apo 
-- PDB ID. 1CFD and the holo -- PDB ID. 1CLL forms) -- as described in
greater detail elsewhere.

 We can check the accuracy of $N_{\mathrm{eff}}$ via a few different
strategies that we have mentioned above -- computing the ESS via the 
correlation times in the structural histograms, and via counting
transitions to and from a known physical state from a dynamic simulation.
Further, we check the robustness of our procedure via different
reference structure for Voronoi bins in determining the
physical states via the method explained in the Appendix (leading to 
possibly slightly different physical states).

 Table~\ref{t2} shows the results for $N_{\mathrm{eff}}$ obtained for
trajectories of butane and calmodulin. Columns 2 to 4 show three
different estimates of $N_{\mathrm{eff}}$ from eq~\ref{e2} using
independent binning, Column 5 shows $N_{\mathrm{eff}}$ obtained
from eq~\ref{e2} using known physical states, 
and Column 6 shows the range of effective
sample size obtained using time correlation analysis. 
For both butane and calmodulin, the procedure is very robust in
estimating $N_{\mathrm{eff}}$, as different binning procedures give
similar estimates. Further, these estimates agree with the range of
sample sizes suggested by the correlation time analysis.
For butane, the first 10 ns of simulation results in 60 transitions
among the three states, and, thus, the whole trajectory (of 
1 $\mathrm{\mu}$s) has about 6000 independent configurations. For
calmodulin, the total number of transitions is 80. These results also
agree with the estimates in Table~\ref{t2}.

 We further use the knowledge of exact physical states in these two
systems to obtain estimates of the sample size using eq~\ref{e2},
instead of using the procedure described in the Appendix to
determine the physical states. For butane, analysis of population
variances in the three known physical states lead to estimating the
sample size as 5865 (as the lowest estimate). A similar
analysis using two physical states of calmodulin leads to
an estimate of $N_{\mathrm{eff}}=91$. Thus, for both these
systems, using known physical states give effective sample sizes
consistent with Table~\ref{t2}.

 One interesting feature emerging from the above analysis is that
the ESS calculated from eq~\ref{e2} is somewhat larger than the
total number of transitions (80) from one state to other in calmodulin.
A possible reason could be that calmodulin is a large flexible
molecule and stays in a particular state
long enough to become decorrelated even before making a transition.

\subsection{Systems with unknown physical states}

 For most biomolecular systems, physical states are not known in advance.
For this reason, we test our method on two systems, leucine dipeptide
(acetaldehyde--(leucine)$_2$--n--methylamide) and
Met--enkaphalin, for which physical states are not known in advance.
We use OPLSAA force field for both systems, and use overdamped
Langevin dynamics (in Tinker v 4.2.2) at 298 K with a friction 
constant of 5/ps for both. For leucine dipeptide we use
an uniform dielectric of 60, and the BS/SA solvation for Met--enkaphalin.
For both the systems, a total of 1 $\mathrm{\mu}$s simulation is
performed with frames stored every 1 ps.

 Again, we estimate $N_{\mathrm{eff}}$ using eq~\ref{e2} using three
different repetitions of the binning procedure, 
and also using time correlation analysis.
The results are shown in Table~\ref{t3}. Again, for both the systems,
we obtain a good agreement with independent binnings to determine
physical states and also with the range obtained using time--correlation 
analysis.

\subsection{Application to discontinuous trajectories}

 In the examples so far to compute ESS, we have used dynamic 
trajectories that
allow for good approximations to the physical states (that are based on
rates of transitions between regions of configuration space). Here, we show
the robustness of our procedure regardless of the actual definitions of
the physical states. 
The motivation behind this exercise is to extend the method efficiently
to non--dynamic trajectories. Although the sample size estimation
using eq~\ref{e2} is applicable to non--dynamic trajectories as well,
the underlying physical states, based on transition rates between regions
of configuration space, cannot be calculated from non--dynamic
trajectories. For such simulations, instead of performing dynamic
simulations as well merely to determine physical states, it is
less computationally intensive to run short dynamic trajectories
starting from configurations already obtained from non--dynamic
simulations.

For this purpose, we divide the dynamic trajectory
into smaller segments of equal length.
We pick only a fraction of configuration at the beginning of each
segment to determine the physical states, as well as population means
and variances. Due to absence of some configurations between two segments,
the trajectories are discontinuous and reflects, conceptually, the
procedure described above for obtaining approximate physical states
for non--dynamic simulations. This could lead to physical states being
determined inaccurately.

However, we find out that as long as each segment is long
enough (with more than 2 independent configurations), we still find
physical states that determine the sample size. We studied the 4 systems and analyzed
the sample sizes by discoutinous trajectories. We took the first 300 ps for each butane segment,
first 1000 Monte Carlo sweeps for each calmodulin segment,
first 1000 ps for each dileucine segment, and first 5000 ps for each met-enkaphalin segment.

Despite the use of discontinuous trajectories to determine the physical
states, we obtain good estimates of the effective sample size in each
case, as shown in Table~\ref{t4} (two estimates using different binnings
are reported).

\subsection{Importance of using physical states}
\label{phys}

 We argued above that population variances in physical states are the most
important descriptors of the sample size. Here we show an example that
quantifies that. The system we use is butane. We divide the 
configurational space into 10 bins using Voronoi cells, and perform
no combination into physical states -- {\it i.e.}, we determine the effective
sample size using these arbitrary bins. For this system, we first performed
the structural histogram analysis discussed in the Introduction to
obtain the correlation time: configurations chosen after this interval
are uncorrelated.  For uncorrelated configurations, we
expect the estimate of the sample size to be independent of particular bins,
however, correlated configurations show a different picture. 

 Table~\ref{t5} shows estimates of the ESS obtained for the 10 arbitrary bins
obtained using uncorrelated and correlated configurations. Clearly, the
correlated trajectory shows a dramatic bin dependence for the estimated
sample size (note that the full correlated trajectory gives consistent
estimates when bins are combined into approximate physical states).

 The difference in bin estimates for correlated configurations 
is because the variance of one bin is a convolution
of state variances and fast processes. To elaborate, we
consider a double well. Imagine that we divide the space into several 
bins, and, say, the seventh of which has the observed probability of $p_{7}$. 
In ideal sampling,
the observed probability should $p_{7}$ in the bin and 1-$p_{7}$
out of the bin. However in dynamic sampling, where
the correlation comes into play, the division of the configuration
space is similar to the separation of the timescales. When the 
system is trapped in state A, with a fractional population $p_{A}$,
its observed probability in the bin turns out to be $p_{7}/p_{A}$
instead of $p_{7}$ and probability out of the bin is 1- $p_{7}/p_{A}$.
The increase of observed probability in the bin comes from correlation.
For example, in MD simulations, the system most likely remains in state A
at the next time increment, unlike ideal sampling.
This dual effects from fast process and state variance leads the new
observed probability.

On the other hand, dynamic correlations identify states -- as discussed
in the Appendix.

\section{Discussion}

 First, we discuss the appropriateness of subdividing a dynamic trajectory
into smaller, equal segments to estimate the population mean and variances.
If the sample size estimate for each of these segments is O(1), then the
method does not reliably give the estimate of the sample size of the
total trajectory, and likely overestimates it. For example, if the correct
total number of independent configurations in the total trajectory is 10,
and we subdivide the whole trajectory into 20 equal segments, then each of the
segment will give a sample size of 1 (minimum number possible) -- leading
to an overestimate of the sample size. In such a scenario, division into
fewer segments is desirable.

 Next, we discuss an appropriate approach in cases that show 
significant asymmetry in state populations.
For example, in a system with two final physical states, separated by a
high barrier, it is possible
that one state has most of the population. The physical states analysis
in the Appendix will correctly give two states, however, the state with
significantly smaller population may not be important for determining
the sample size. In our examples, we found that if a state has
less than 5$\%$ of the population, then it is not meaningful for
determining the sample size.

\section{Conclusions}

We have developed a new method to assess the quality of molecular
simulation trajectories -- effective sample size -- using variances
of population in the physical states. A major feature of
this method is that it is applicable both to dynamic and non--dynamic 
methods, and gives a tool to compare sampling and efficiencies of 
different molecular simulation algorithms. Another feature of
this procedure is that it is applicable to discontinuous trajectories
as well. We also demonstrated that
our procedure is robust with respect to actual definition of
physical states -- a fact that is expected to be of importance
for systems for which actual physical states are not known in
advance.

To supplement the estimation of the effective sample size, we also
developed a procedure for automated determination of physical
states. This procedure yields, in a natural way, a heirarchical picture of the
configurational space, based on transition rates between regions of 
configurational space.

\clearpage

\section*{Appendix: Physical states determination from dynamic trajectories}

In this Appendix, we give a brief description on our physical state
discovery method and its results. 
In this method, bins in configurational space are combined to
give the physical states, as discussed below in more detail.
There is no Markovian requirement on the selection of bins. Indeed,
a typical bin in a configurational space for a large multidimensional
system may itself encompass several separate minima.

 Our method is based on the work
of Chodera {\it et al.}\cite{chodera}, but is simpler. In addition, the 
following
method clearly shows the heirarchical nature of the configurational
space, and focuses on the slowest timescale -- which is of
paramount importance for the estimation of the effective sample size
in the main text.

\section*{Method}

\subsection*{Use of rates to describe conformational dynamics}

Rates between regions of configuration space are a fundamental property
that emerges uniquely from the natural system dynamics. We use rates
between regions of configurational space to determine the physical
states. We first decompose the conformational space into multiple
bins. Subsequently, we combine some of the bins that have high transition
probability between them. This procedure is based on the physical
idea of separation of time scales: there is a fast timescale (high
transition rates) associated with regions within a single physical
state, and a slow time scale (low transition rates) associated with
regions of configurational space separated by high barriers. Thus,
transition rates between regions of configurational space form a fundamental
basis for determination of physical states.

\subsection*{Binning decomposition of the configurational space}

We divide the whole configuration space into a large number of bins,
and determine the physical states by combination of these regions.
The procedure to decompose the whole configurational space (with $N$
configurations) into $M$ bins is described as follows: 
\begin{enumerate}
\item A configuration $i$ is picked at random from the trajectory. 
\item The distance, by use of an appropriate metric as discussed later,
of the configuration $i$ to to all remaining configurations in the
trajectory is, then, computed. 
\item The configurations are rearranged by the distance in an ascending
order. The first $1/M\times N$ configurations are removed. 
\item Steps 1--3 are repeated $M-1$ times on progressively smaller set
of remaining configurations, resulting in a total of $M$ reference
configurations, $i$. 
\end{enumerate}
For the distance metric, we select the root-mean squared deviation
(RMSD) of the full molecule, estimated after alignment. In contrast,
using just the backbone RMSD may be a poor distance metric as it ignores
potentially relevant side chain kinetics.

After references are selected, we decompose the whole space into bins
based on a Voronoi construction. For each configuration, we calculate
the RMSD of this configuration from each reference structure. We assign
this configuration to the bin associated with the reference structure,
$i$, with which the configurations has the smallest RMSD.

\subsection*{Calculation of rates among bins and bin combination}

We compute the mean first passage time (MFPT) from each bin, $i$,
to every other bin, $j$, using the full dynamic trajectory. The rate
from bin $i$ to bin $j$ is the inverse of that MFPT. In general,
the rate from bin $i$ to bin $j$ is not the same as the rate from
bin $j$ to bin $i$ -- and we take a linear average of these two
rates to define the rate between bin $i$ and bin $j$. Subsequently,
we chose a cutoff rate, $k_{c}$, and combine all pairs of bins that
show a rate higher than $k_{c}$.

\subsection*{Hierarchy}

An obvious manner in which the hierarchical picture emerges from the
above discussion is from the choice of $k_{c}$. An increase in $k_{c}$
leads to the combination of fewer bins -- resulting in an increase
in the final number of states obtained. Starting from a low value
of $k_{c}$, we obtain a hierarchical picture with increasing number
of states upon increasing $k_{c}$.

This hierarchical picture can be significantly affected by the time
interval underlying the MFPT calculations. For example, although a
trajectory may have a low likelihood (hence a low rate) to cross over
the $2kT$ barrier in Figure~1 in time $\tau_{1}$, it may easily
cross that barrier for a long enough time interval, $\tau_{2}$. Thus,
a hierarchical picture at the lowest level can differentiate the two
left states of Figure~1 if the rates are computed from the dynamic
trajectory at every $\tau_{1}$ interval. On the other hand, if the
rates are computed using the $\tau_{2}$ interval, $2kT$ barrier
cannot be resolved at the lowest hierarchical level. As an extreme
case, if $\tau$ is chosen to be longer than the largest decorrelation
time in the system, then the rates to a bin from any other bin is
simply proportional to the equilibrium population of that particular
bin -- the application of the procedure described above is not appropriate.

\section*{Results}

\begin{figure}
\begin{center}
\resizebox{6in}{!}{\includegraphics{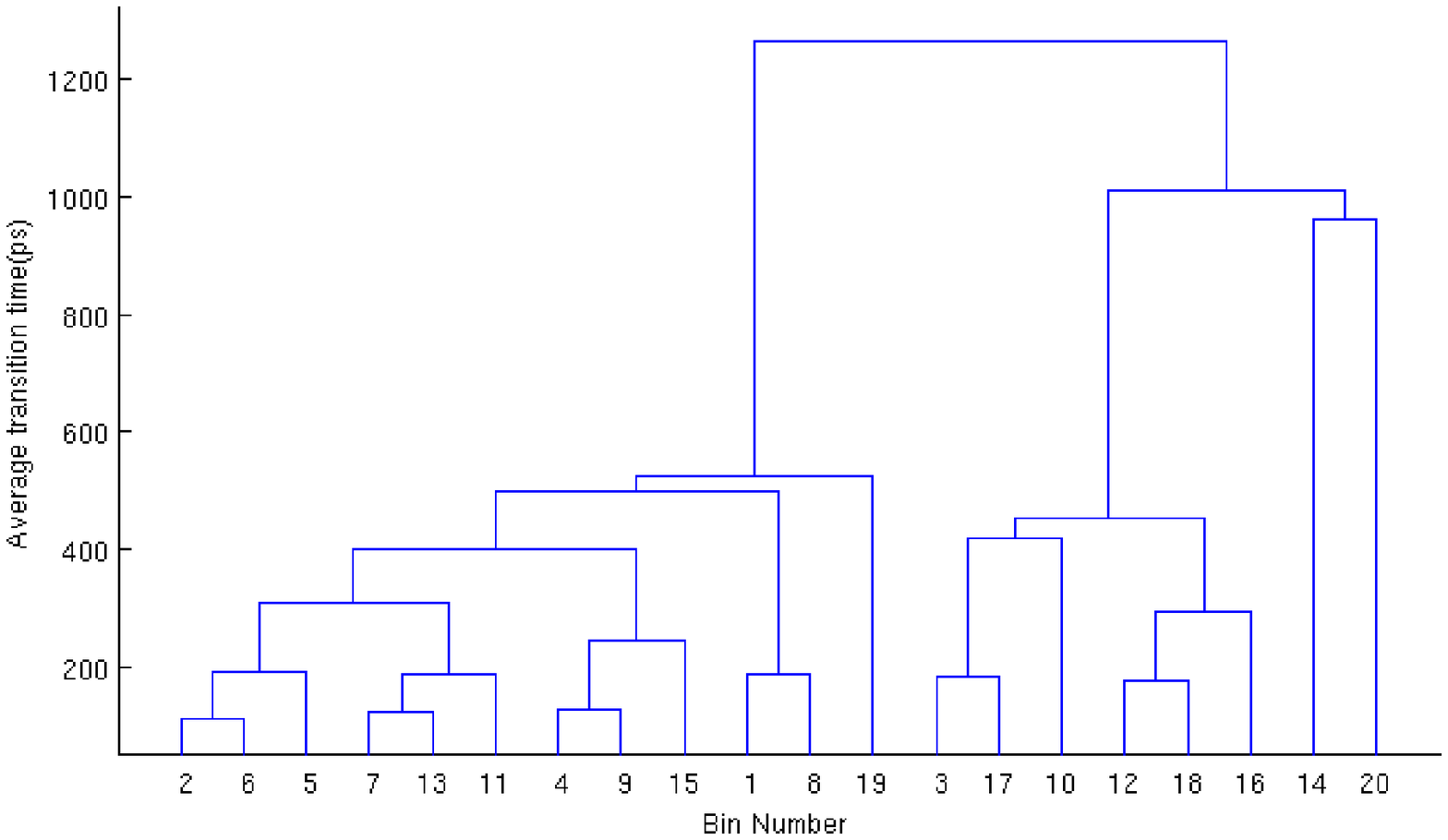}}
\caption{Hierarchical physical states for dileucine shown via the
average transition time required for transition among bin pairs.
Bin pairs that combine ``faster'' ({\it i.e.}, have shorter transition
time) are combined at a lower level of the heirarchy.}
\label{dil}
\end{center}
\end{figure}

\begin{figure}
\begin{center}
\resizebox{6in}{!}{\includegraphics{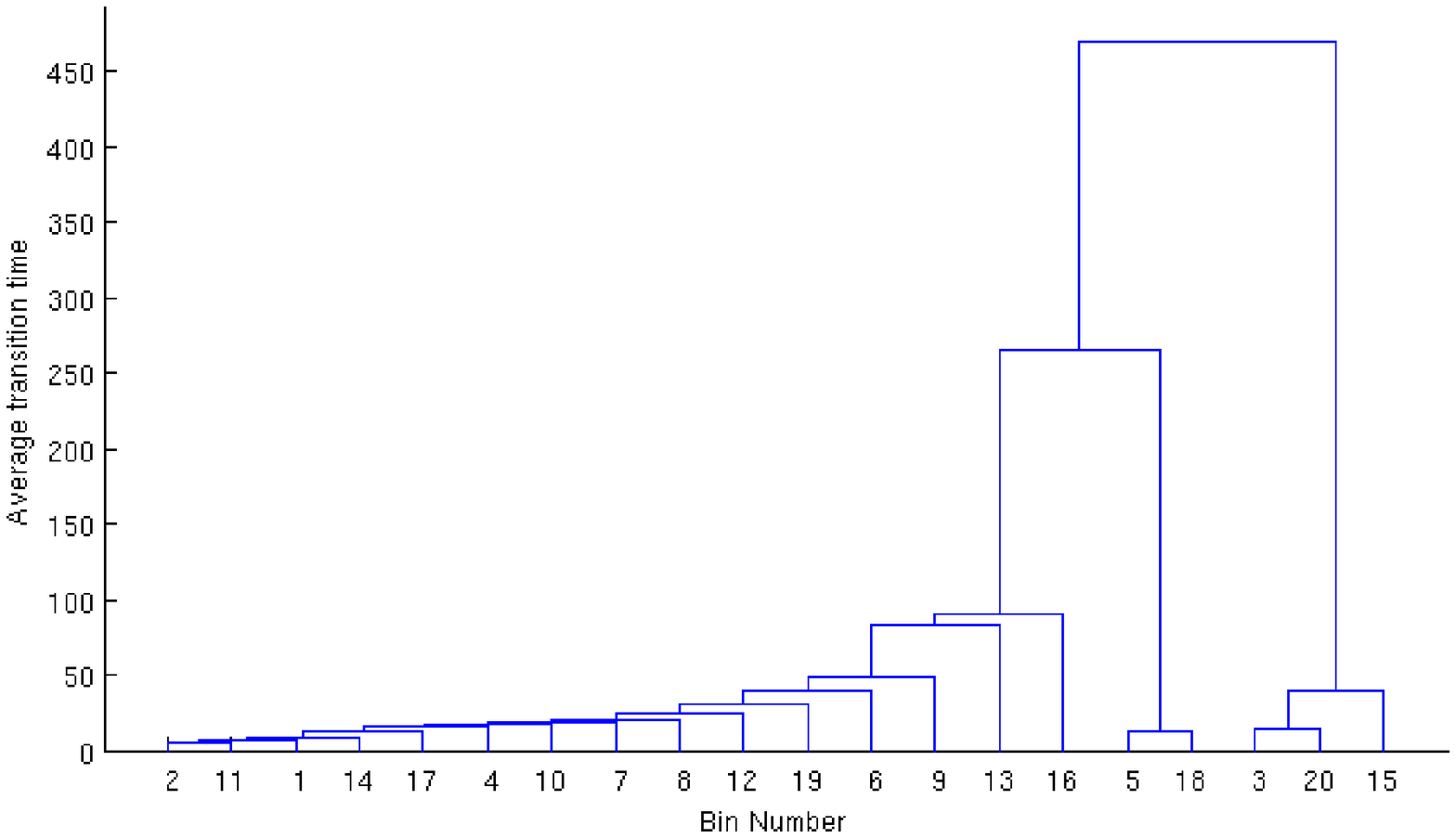}}
\caption{Hierarchical physical states for butane shown via the
average transition time required for transition among bin pairs.
Bin pairs that combine ``faster'' ({\it i.e.}, have shorter transition
time) are combined at a lower level of the heirarchy.}
\label{butane}
\end{center}
\end{figure}

Figures~\ref{dil} and \ref{butane} show the hierarchical physical for
dileucine and butane, respectively. They both start from 20 bins and combine 
all the way
to single states. We took the second last step (2 states) to calculate
sample size. Dileucine is very flexible systems, due to which
the transition time goes up smoothly, while butane get 3 states at
early stage and have a sharp increase in transtion time among those
3 states.

\clearpage

\begin{table}
\begin{center}
\caption{Effective sample sizes for butane and calmodulin using eq~\ref{e2}
and three different reptitions of the binning procedure in Columns 2--4, 
using eq~\ref{e2} and known physical states in Column 5, and
as obtained by the structural time correlation in Column 6.}
\begin{tabular}{c|c|c|c|c|c}
System & eq~\ref{e2} (1)
& eq~\ref{e2} (2)
& eq~\ref{e2} (3)
& known physical states
& Time correlation function \\ \hline
butane & 6064 & 6236 & 6200 & 5865 & 5000--10000 \\
calmodulin & 93 & 90 & 92 & 91 & 80--160 \\
\end{tabular}
\label{t2}
\end{center}
\end{table}

\clearpage

\begin{table}
\begin{center}
\caption{Effective sample sizes for di--leucine and Met--enkaphalin 
using eq~\ref{e2}
and three different reptitions of the binning procedure in Columns 2--4, and
as obtained by the structural time correlation.}
\begin{tabular}{c|c|c|c|c}
System & eq~\ref{e2} (1)
& eq~\ref{e2} (2)
& eq~\ref{e2} (3)
& Time correlation function \\ \hline
di--leucine & 1982 & 1878 & 1904 & 1100-2200 \\
Met--enkaphalin & 416 & 362 & 365 & 250--500 \\
\end{tabular}
\label{t3}
\end{center}
\end{table}

\clearpage

\begin{table}
\begin{center}
\caption{Effective sample sizes for discontinuous trajectories obtained
from eq~\ref{e2} by using two reptitions of the binning procedure.}
\begin{tabular}{c|c|c}
System & eq~\ref{e2} (1)
& eq~\ref{e2} (2) \\ \hline
butane & 5842 & 6024 \\
calmodulin & 91 & 92 \\
dileucine & 1803 & 2016 \\
Met--enkaphalin & 397 & 336 \\
\end{tabular}
\label{t4}
\end{center}
\end{table}

\clearpage

\begin{table}
\begin{center}
\caption{Butane sample size estimated in 10 different bins using
correlated and uncorrelated sampling. The actual sample size is 2000.}
\begin{tabular}{c|c|c}
Bin number  &  uncorrelated & correlated \\ \hline
1  & 1882 & 1256 \\
2  & 2415 & 61380 \\
3  & 1837 & 82080 \\
4  & 2444 & 91820 \\
5  & 1866 & 292640 \\
6  & 2172 & 71180 \\
7  & 1892 & 240200 \\
8  & 2264 & 5600 \\
9  & 3936 & 162720 \\
10  & 2040 & 310260 \\
\end{tabular}
\label{t5}
\end{center}
\end{table}

\clearpage

\clearpage

\clearpage


\begin{thebibliography}{10}

\bibitem{frenkel_smit}
Frenkel, D \& Smit, B.
\newblock (2002) {\em Understanding Molecular Simulations}.
\newblock (Academic Press, San Diego).

\bibitem{ergodic1}
Mountain, R.~D \& Thirumalai, D.
\newblock (1989) {\em J. Phys. Chem.} {\bf 93}, 6975--6979.

\bibitem{ergodic2}
Mountain, R.~D \& Thirumalai, D.
\newblock (1990) {\em Int. J. Mod. Phys. C} {\bf 1}, 77--89.

\bibitem{block}
Flyvbjerg, H \& h.~G.~Petersen.
\newblock (1989) {\em J. Chem. Phys.} {\bf 91}, 461--466.

\bibitem{ed2}
Lyman, E \& Zuckerman, D.~M.
\newblock (2007) {\em J. Phys. Chem. B} {\bf 111}, 12876--12882.

\bibitem{md}
Adler, B.~J \& Wainwright, T.~E.
\newblock (1959) {\em J. Chem. Phys.} {\bf 31}, 459--466.

\bibitem{metropolis}
Metropolis, N, Rosenbluth, A.~W, Rosenbluth, M.~N, Teller, A.~H,  \& Teller, E.
\newblock (1953) {\em J. Chem. Phys.} {\bf 21}, 1087--1092.

\bibitem{rex1}
Swendsen, R.~H \& Wang, J.~S.
\newblock (1986) {\em Phys. Rev. Lett} {\bf 57}, 2607--2609.

\bibitem{rex2}
Earl, D.~J \& Deem, M.~W.
\newblock (2005) {\em Phys. Chem. Chem. Phys.} {\bf 7}, 3910.

\bibitem{rex3}
Hansmann, U. H.~E.
\newblock (1997) {\em Chem. Phys. Lett.} {\bf 281}, 140--150.

\bibitem{rosenbluth}
Rosenbluth, M.~N \& Rosenbluth, A.~W.
\newblock (1955) {\em J. Chem. Phys.} {\bf 23}, 356--359.

\bibitem{wall}
Wall, F.~T \& Erpenbeck, J.~J.
\newblock (1959) {\em J. Chem. Phys.} {\bf 30}, 634--637.

\bibitem{grassberger}
Grassberger, P.
\newblock (1997) {\em Phys. Rev. E} {\bf 56}, 3682--3693.

\bibitem{liu}
Liu, J.~S.
\newblock (2004) {\em Monte Carlo Strategies in Scientific Computing}.
\newblock (Springer, New York).

\bibitem{chodera}
Chodera, J.~D, Singhal, N, Swope, W.~C, Pande, V.~S,  \& Dill, K.~A.
\newblock (2007) {\em J. Chem. Phys.} {\bf 126}, 155101.

\end{thebibliography}
\end{document}